\documentstyle[12pt,axodraw]{article}
\begin{document}
\baselineskip 0.6cm
\newcommand{\gsim}{ \mathop{}_{\textstyle \sim}^{\textstyle >} }
\newcommand{\lsim}{ \mathop{}_{\textstyle \sim}^{\textstyle <} }


\begin{titlepage}

\begin{flushright}
UCB-PTH-01/08 \\
LBNL-47610 \\
\end{flushright}

\vskip 0.5cm

\begin{center}
{\Large \bf  Gauge Unification in Higher Dimensions}

\vskip 1.0cm

{\large 
Lawrence Hall and Yasunori Nomura
}

\vskip 0.5cm
 {\it Department of Physics, \\ and \\
Theoretical Physics Group, Lawrence Berkeley National Laboratory,\\
University of California, Berkeley, CA 94720}

\vskip 1.0cm

\abstract{
A complete 5-dimensional $SU(5)$ unified theory is constructed 
which, on compactification on the orbifold with two different 
$Z_2$'s ($Z_2$ and $Z_2'$), yields the minimal 
supersymmetric standard model. The orbifold accomplishes $SU(5)$ 
gauge symmetry breaking, doublet-triplet splitting, and a vanishing 
of proton decay from operators of dimension 5. Until 4d supersymmetry 
is broken, all proton decay from dimension 4 and dimension 
5 operators is forced to vanish by an exact $U(1)_R$ symmetry. 
Quarks and leptons and their Yukawa interactions are 
located at the $Z_2$ orbifold fixed points, where $SU(5)$ is 
unbroken. A new mechanism for introducing $SU(5)$ breaking into 
the quark and lepton masses is introduced, which originates from  
the $SU(5)$ violation in the zero-mode structure of bulk multiplets. 
Even though $SU(5)$ is absent at the $Z_2'$ orbifold fixed point,
the brane threshold corrections to gauge coupling unification are 
argued to be negligibly small, while the logarithmic corrections 
are small and in a direction which improves the agreement with the 
experimental measurements of the gauge couplings. Furthermore, the 
$X$ gauge boson mass is lowered, so that $p \rightarrow e^+ \pi^0$ 
is expected with a rate within about one order of magnitude of the 
current limit. Supersymmetry breaking occurs on the $Z_2'$ orbifold 
fixed point, and is felt directly by the gauge and Higgs sectors, 
while squarks and sleptons acquire mass via gaugino mediation, 
solving the supersymmetric flavor problem.}

\end{center}
\end{titlepage}


\section{Introduction}
\label{section:intro}

The successful prediction of the weak mixing angle is a major
achievement of particle theory of recent decades \cite{Georgi:1974yf}. 
It suggests that weak-scale supersymmetry with two light Higgs
doublets will soon be discovered at colliders, and
there should be no exotic states in an energy desert up to 
$M_U \approx 10^{16}$ GeV. The physics immediately above this 
unification scale has been viewed in three frameworks: 
4 dimensional grand unified field theories \cite{Georgi:1974sy}, 
higher-dimensional grand unified theories motivated by string 
theory \cite{Candelas:1985en}, and string theory \cite{Horava:1996qa}. 
Grand unification in 4d provides a simple 
and elegant understanding of the quantum numbers of the quarks and 
leptons in a generation, but these successes are open to doubt because 
of several issues which require considerable effort to overcome. Chief 
amongst these are
\begin{itemize}
\item How is the grand unified gauge symmetry to be broken?
\item Why are the weak doublet Higgs bosons split in mass from the
colored triplet partners?
\item Why have we not already observed proton decay induced by
dimension 5 operators?
\item Why should the two Higgs doublets be light when the standard model
gauge symmetry allows a large mass term?
\end{itemize}

In this paper we study higher-dimensional grand unified field 
theories above $M_U$, without inputing suggestions from string theory, 
for example on the number of extra dimensions and the gauge group. 
We follow a ``bottom-up'' approach, seeking simple solutions 
for the above issues.  Many of the methods we use, for example 
for gauge symmetry breaking and doublet-triplet splitting, were 
introduced in the string-motivated context \cite{Candelas:1985en}. 
Recently, Kawamura has shown how the first two issues highlighted above 
are simply and elegantly solved in the case of an $SU(5)$ unified gauge 
symmetry \cite{Kawamura:2000ev} using the orbifold 
$S^1/(Z_2 \times Z_2')$ previously introduced for breaking 
weak-scale supersymmetry \cite{Barbieri:2000vh}. 
Orbifolding along one axis, using a parity which is $+$ for the
weak direction but $-$ for the strong direction, automatically breaks
$SU(5)$ to the standard model gauge group, and gives
zero modes for the Higgs doublets but not for the triplet partners.
There is no need for Higgs multiplets at the unified scale with a scalar
potential designed to give the correct unified symmetry breaking, and there
is no need to arrange couplings in such a way that only the Higgs
triplets acquire mass. An important recent observation is that to
proceed further with construction of such theories it is necessary to
consider how the quarks and leptons transform under the $Z_2 \times
Z_2'$ symmetry \cite{Altarelli:2001qj}. However, the brane
interactions advocated explicitly break 
the 5-dimensional $SU(5)$ gauge symmetry.
In this paper we require that all interactions preserve 
this symmetry.

In this paper we construct completely realistic unified theories based 
on the orbifold $S^1/(Z_2 \times Z_2')$. 
We show that the answer to the first question is that, from the 5d 
viewpoint, the unified gauge symmetry is unbroken but takes a restricted 
form, while from the 4d viewpoint there is no unbroken Georgi-Glashow 
$SU(5)$.  We show that higher dimensional unified theories
generically do not have proton decay from dimension 5 operators. In
particular the usual dimension 5 proton decay, resulting from the
exchange of the colored triplet Higgs multiplet \cite{Sakai:1982pk},
is absent because a $U(1)_R$ symmetry forces a special form for the 
masses of these states. Furthermore, this $U(1)_R$ symmetry forbids the
appearance of dimension 5 operators which violate baryon number in the
superpotential. This also allows a fundamental distinction between
Higgs and matter fields, and provides an understanding of why the Higgs
doublet multiplets are light even though they are vectorial under the
standard model: they are protected from a mass term by the same bulk
$U(1)_R$ symmetry that forbids proton decay at dimension 5. Thus we
see that the four issues of grand unification listed above are
automatically solved in higher dimensional theories. The framework for 
such 5 dimensional theories compactified on a  $S^1/(Z_2 \times Z_2')$ 
orbifold is given in section \ref{section:orbifold}, together with 
an elucidation of some of their general properties.
A complete theory is given with gauge group $SU(5)$ in section 
\ref{section:su5}.

In addition to these accomplishments, the complete theory constructed in
section \ref{section:su5} illustrates several new results:
\begin{itemize}
\item Unified quark and lepton  multiplets, and Yukawa couplings 
on orbifold fixed points, can be made consistent with the 
orbifolding parity which restricts the unified symmetry.
\item Even though the unified gauge symmetry is restricted at orbifold
fixed points, the threshold corrections to the weak mixing angle, both
at the classical and quantum level, are small, preserving the
successful prediction for the weak mixing angle.
\item The orbifold construction for the unified gauge symmetry
breaking provides a natural location for gaugino mediated supersymmetry 
breaking \cite{Kaplan:2000ac} --- all the ingredients needed for gaugino
mediation are automatically present in the theory.
\item Even though the theory possesses an $SU(5)$ gauge
symmetry which, from the higher dimensional viewpoint is unbroken, 
the Yukawa couplings,
which respect this symmetry, need not have the usual $SU(5)$ relations
amongst the massless 4d states.
\end{itemize}
The first two points are crucial if these theories are to predict
$\sin^2 \theta_w$ and provide an understanding of the quark and lepton
gauge quantum numbers. The last
two point allows us to construct completely realistic theories.

\section{Unified Gauge Symmetry Transformations \\ on Orbifold Spacetime}
\label{section:orbifold}

In this section we introduce a class of higher dimensional unified
field theories, concentrating on the nature of the gauge symmetry 
transformation. We then discuss several features which are generic 
to this class.
For simplicity we consider a single compact extra dimension, $y$ 
($=x_5$), and assume a fixed radius with size given by the 
unification scale.  We take the unified
gauge interactions in 5 dimensions to have gauge group $G$. 
The Higgs doublets also propagate in 5 dimensions as components 
of hypermultiplets. Using 4 dimensional
superfield notation, we write the vector multiplet as $(V,\Sigma)$,
where $V$ is a 4d vector multiplet and $\Sigma$ a chiral adjoint, and
the hypermultiplet as $(H,H^c)$, where $H$ and $H^c$ are chiral
multiplets with opposite gauge transformations. 

The form of the gauge transformations under $G$ can be restricted by 
compactifying on the orbifold $S^1/Z_2$ with a 
parity, $P$, which acts on the vector representation of $G$, making some
components positive and some components negative: $P=(+,+,... -,-,...)$.
The orbifold symmetry on any tensor field $\phi$ is then defined by
$\phi(-y)  = P \phi(y)$, where $P$ acts separately on all vector
indices of $\phi$, and an overall sign choice for the parity of the
multiplet may also be included in $P$. 
It is understood that there is a relative minus sign 
between the transformation of $H^c(\Sigma)$ and $H (V)$, as required 
by the $P$ invariance of the 5d gauge interactions.
In certain cases $P$ is a discrete gauge transformation, but it need
not be. 

A non-supersymmetric theory with $G=SU(5)$ and $P=(-,-,-,+,+)$ was
considered by Kawamura \cite{Kawamura:2000nj}; here we discuss a
supersymmetric version.  The Higgs bosons are taken to lie 
in a hypermultiplet $(H,H^c)(x,y)$, with $H$ and $H^c$ chiral multiplets
transforming as ${\bf 5}$ and $\bar{\bf 5}$ representations. The orbifold
projection accomplishes doublet-triplet splitting, in the sense that
$H$ has a weak doublet zero mode but not a color triplet zero
mode. However, the projections work oppositely for $H^c$ which
contains only a color triplet zero mode. Similarly, while the $X$ gauge 
bosons have negative $P$ and therefore no zero mode, the exotic color
triplet, weak doublet states in the chiral adjoint, $\Sigma_X$, does
have a zero mode. Such exotic light states are generic to orbifolding 
with a single $Z_2$ and lead to an incorrect prediction 
for the weak mixing angle.  These exotic states can be removed 
by introducing two sets of different orbifold parities, giving 
additional structures to the spacetime, which we study in the rest of 
this paper.

\begin{figure}
\begin{center} 
\begin{picture}(160,180)(-80,-90)
  \CArc(0,0)(60,0,360)
  \DashLine(-80,0)(80,0){2} \Text(-90,0)[r]{$P$}
  \DashLine(0,-80)(0,80){2} \Text(0,90)[b]{$P'$}
  \CArc(0,0)(61,180,205) \CArc(0,0)(59,180,205)
  \Line(-54,-26)(-52,-20) \Line(-54,-26)(-60,-23) \Text(-48,-22)[l]{$y$}
  \CArc(0,0)(61,270,295) \CArc(0,0)(59,270,295)
  \Line(26,-54)(20,-52) \Line(26,-54)(23,-60) \Text(25,-46)[b]{$y'$}

  \Vertex(-60,0){3} \Text(-65,10)[br]{$O$}
  \Vertex(0,-60){3} \Text(-5,-70)[ur]{$O'$}
  \Vertex(60,0){3}  \Text(65,-10)[ul]{$O$}
  \Vertex(0,60){3}  \Text(5,70)[bl]{$O'$}
\end{picture}
\caption{$S^1/(Z_2 \times Z_2')$ orbifold in the fifth dimension.}
\label{Fig_space}
\end{center}
\end{figure}
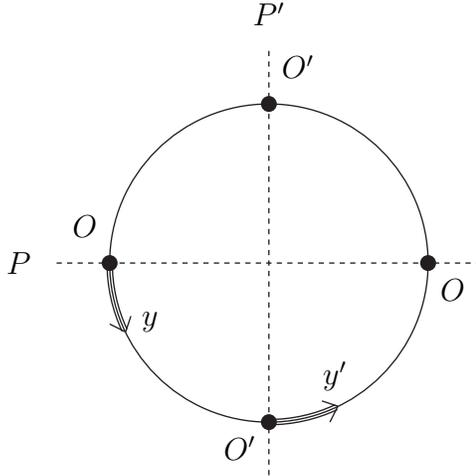
We choose to view this spacetime as illustrated in Figure 
\ref{Fig_space}. There are two reflection symmetries, 
$y \rightarrow -y$ and  $y' \rightarrow -y'$, each with its 
own orbifold parity, $P$ and $P'$, acting on the fields. 
Each reflection introduces special points, $O$ and $O'$, which are 
fixed points of the transformation.  The physical space can be taken 
to be $0 \leq y \leq \pi R/2$, and is the usual $S^1/Z_2$ orbifold.
Nevertheless, we find it convenient to label the pattern in 
Figure \ref{Fig_space} as $S^1/(Z_2 \times Z_2')$.
The components of the vector multiplet can be assembled into four 
groups, $V_{PP'}$, according to their transformation properties. 
They have Kaluza-Klein (KK) mode expansions as given in 
Eqs.~(\ref{eq:expansion-1}--\ref{eq:expansion-4}), with $\phi \rightarrow V$. 
Before orbifolding the gauge transformations are arbitrary
functions of position $\xi = \xi(y)$ for each generator (we suppress
the dependence on the usual four dimensions). However, the non-trivial
orbifold quantum numbers of the gauge particles imply that the most
general set of gauge transformations are restricted to have the form
\begin{equation}
  U = \exp \left[ i \left(\xi_{++}(y) T_{++} + \xi_{+-}(y) T_{+-} + \xi_{-+}(y)
  T_{-+} + \xi_{--}(y) T_{--} \right) \right].
\label{eq:gauge-transformation}
\end{equation}
The generators of the gauge group $T$ are labelled by the $P$ and $P'$
quantum numbers of the corresponding gauge boson. The gauge functions
$\xi_{PP'}(y)$ also have the KK mode expansions similar to those of 
Eqs.~(\ref{eq:expansion-1}--\ref{eq:expansion-4}),
with $\phi \rightarrow \xi$. At an arbitrary point in the bulk all
generators of the gauge group are operative. However, at the fixed
points $O$ the gauge transformations generated by $T_{-P'}$
vanish, because $\xi_{-P'}(y=0,\pi R) =0$. 
Thus locally at these points the symmetry should be thought of
as the subgroup $H$ generated by the set of generators
$T_{+P'}$.\footnote{
If interactions on $O$ contain $\Sigma$ fields, there are 
additional constraints on the form of the interactions coming from 
$\partial_y \xi_{-P'}(y=0,\pi R) \neq 0$.}
Similarly for the points $O'$; hence
\begin{equation}
  G \stackrel{O}  {\rightarrow} H, \;\;\;\;\;\;\;\;\;\; 
  G \stackrel{O'} {\rightarrow} H'.
\end{equation}
Relative to the theory with arbitrary $G$ gauge transformations on a
circle, one can view the orbifold procedure as leading to a local
explicit breaking of some of these gauge symmetries at the fixed points. 
In the model of the next section, $H'$ is the standard model gauge group, 
so that $O'$ is simply not affected by the unified gauge transformations. 
From the 5d viewpoint, working on the orbifold spacetime, the gauge
symmetry simply takes a restricted form --- the theory does not
involve any gauge symmetry breaking.  

Once a KK mode expansion is made, one finds that the 5d gauge 
transformation corresponds to infinite towers of gauge transformations, 
which mix up KK modes of different levels, associated with the KK gauge 
boson modes $A_\mu^{(n)}$.  In this 4d picture, $\partial_y$ acts like a 
symmetry breaking vacuum expectation value and mixes $A_\mu^{(n)}$ with 
$\partial_\mu A_5^{(n)}$, giving a mass to the $n\neq 0$ modes.  
Without an orbifold procedure, the theory possesses an infinite number of 
gauge transformations parametrized by an integer $n$ for each generator.  
However, after the orbifolding, some of the gauge transformations are 
projected out due to the non-trivial orbifold quantum numbers of the 
gauge parameters, and $n$ can no longer take arbitrary values.
Therefore, from a 4d viewpoint, the orbifold procedure corresponds to 
imposing only the restricted sets of gauge transformations on the theory.  
The unbroken gauge symmetry of the 4d theory is generated by $T_{++}$, 
and is the intersection of $H$ and $H'$. 

Our class of theories includes many possibilities for $G$, $P$ and $P'$. We
require that the choice of $G$ and $P$ is such that $H$, the group of
local gauge transformations at $O$, is, or contains, $SU(5)$. To
obtain an understanding of the quark and lepton quantum numbers we
require that the three generations are described by brane fields at $O$
in representations of $H$: $3( T_{\bf 10} + F_{\bar{\bf 5}})$ for $H=SU(5)$. 
Furthermore, Yukawa couplings of this matter to the bulk Higgs
field will be placed at $O$. Notice that
the standard model quarks, leptons, gauge fields and Higgs bosons are
transforming under a unified symmetry group, such as $SU(5)$, which
from the 5d viewpoint, is
{\it unbroken}. How can this be possible? The answer is that this is
not the same 4d $SU(5)$ symmetry that Georgi and Glashow introduced. 
While it acts
in a standard fashion on the quarks and leptons, its action on the
gauge and Higgs fields, which live in the bulk is non-standard. For
example, consider the gauge transformation induced by the generator
$T_X$, which is in $SU(5)/(SU(3)_C \times SU(2)_L \times U(1)_Y)$, assuming
that $P'$ has been chosen such that $T_X = T_{X+-}$. This assignment
guarantees that if the Higgs doublets are $h_{2++}(y)$, as they should
be to possess zero modes, then the color triplet partners will be
$h_{3+-}(y)$, and will not have zero modes. Making a KK mode expansion,
one discovers that the $T_X$ transformation
rotates the zero-mode $h_2^{(0)}$ into a combination of massive triplet
KK modes $h_3^{(n)}$, and, from the 4d viewpoint, this gauge transformation 
is broken. It is because $\xi_X = \xi_{X+-}$ has no zero-mode 
contribution that there is no 4d Georgi-Glashow $SU(5)$ symmetry.
A central thesis
of this paper is that the grand unified symmetry is an inherently 
extra-dimensional symmetry --- it is not a 4d symmetry which must be
spontaneously broken, and it does not contain the 4d Georgi-Glashow
$SU(5)$ as a subgroup.

The construction of theories of the type outlined above is not
guaranteed. There are certain consistency conditions that must be
imposed. All interactions, both bulk and brane, must be invariant
under each parity, that is under $y \rightarrow -y$ and $\phi(-y)
\rightarrow P \phi(y)$, and similarly for $P'$. For the bulk gauge
interactions this means that the group structure constants must 
obey $f^{\hat{a}\hat{b}\hat{c}} = f^{\hat{a}bc} =0$, where
$a(\hat{a})$ runs over generators even (odd) under $P$. Important
constraints result because the brane interactions at $O$, which
include the Yukawa couplings for the quarks and leptons, must be
symmetrical under both $P'$ and the gauge symmetry $H$. Constructing
such interactions is non-trivial, and will be discussed in detail in
the explicit theory of section \ref{section:su5}.

The parity $P'$ must be chosen so that $h_3$ and $V_X$ are odd and
have no zero modes. This implies that the gauge symmetry $H'$ at the fixed 
points $O'$ does not contain $SU(5)$. Hence brane interactions at $O'$
violate the Georgi-Glashow $SU(5)$ symmetry. This raises the crucial
question as to whether the prediction for the weak mixing angle
survives in these theories --- is Georgi-Glashow $SU(5)$ necessary for
a successful prediction of the weak mixing angle? We find that it is
not: the three standard model gauge interactions are unified into a
single non-Abelian higher dimensional gauge symmetry, so that a zero-th
order relation amongst the three gauge couplings is preserved.
Threshold corrections have two origins: 
brane kinetic terms for gauge fields located at $O'$, and
$SU(5)$ splitting of the KK multiplets of the Higgs and gauge towers,
which are inherent to our class of theories. In the
next section we show that these threshold corrections are under control 
and small. It is remarkable and non-trivial that, even if the brane 
gauge kinetic terms violate Georgi-Glashow $SU(5)$ strongly, the 
resulting tree-level corrections to the weak mixing angle are negligible.

Proton decay from color triplet Higgsino exchange vanishes in our
class of theories. These Higgsinos acquire mass via the KK mode
expansion of operators of the form $H \partial_y H^c$. 
The Dirac mass couples the triplet Higgsino to a
state in $H^c$ which does not couple to quarks and
leptons. It is well known that the dimension 5 proton decay problem
can be solved by the form of the triplet Higgsino mass matrix; in
higher dimensional unification the structure of the theory requires
this dimension 5 proton decay to be absent. The form of the mass
matrix is guaranteed by a $U(1)_R$ symmetry of the 5d gauge
interactions. As shown in the next section, this can be trivially 
extended to the brane matter fields so that all proton decay at
dimension 5 is absent, and $R$ parity automatically results, giving
proton stability also at dimension 4.

Since matter resides at $O$, where the gauge transformations include
those of $SU(5)$, the Yukawa interactions are necessarily $SU(5)$
symmetric, leading to the unified fermion mass relation $m_s/m_d =
m_\mu/m_e$, which conflicts with data by an order of magnitude. Does
this exclude our framework? In 4d unified theories, acceptable mass
relations follow from using higher dimensional operators involving
factors of $SU(5)$ symmetry breaking vacuum expectation values, 
$\langle \Sigma_{SU(5)} \rangle$. Indeed the Yukawa matrices can be 
viewed as expansions in $\langle \Sigma_{SU(5)} \rangle / M_{\rm Pl}$
\cite{Ellis:1979fg}, allowing the construction of predictive unified 
theories of fermion masses \cite{Georgi:1979df, Dimopoulos:1992yz}. 
This option is not available to us as there are no $SU(5)$ breaking
vacuum expectation values. We find that our framework leads 
to an alternative origin for $SU(5)$ breaking 
Yukawa interactions, and we are able to introduce a
new, highly constrained, class of predictive unified theories of
fermion masses. The only origin of $SU(5)$ breaking at $O$ is through
the zero-mode structure of bulk fields with non-zero wavefunctions at
$O$. We therefore introduce a new class of bulk matter in a vector
representation of $G$ as hypermultiplets $(B,B^c) + (\bar{B},
\bar{B}^c)$. The symmetry quantum numbers of these fields are assigned
such that brane interactions at $O$ give mass terms
and Yukawa interactions which involve both the brane matter and the
bulk matter. For example, when $G$ is $SU(5)$, and $B$ is taken to be
a 5-plet, the brane interactions at $O$ include the mass terms
$B(\bar{B} + F_{\bar{\bf 5}})$ and the Yukawa interactions $T_{\bf 10} 
(\bar{B} + F_{\bar{\bf 5}}) H_{\bar{\bf 5}}$. This leads to realistic 
masses, as discussed in the next section.

Below the compactification scale our theory is the minimal
supersymmetric standard model. While there are a variety of
possibilities for supersymmetry breaking, it is readily apparent that
our class of theories provides a very natural setting for gaugino 
mediation. Gaugino mediation \cite{Kaplan:2000ac} requires the
standard model gauge fields
to propagate in a bulk which contains at least two
branes. One brane has quarks and leptons localized to it, while the
other is the location of supersymmetry breaking. 
Clearly, in our class of theories the
supersymmetry breaking should reside at $O'$.

\section{An $SU(5)$ Theory}
\label{section:su5}

\subsection{Orbifold symmetry structure}

In this section we construct a complete unified $SU(5)$ theory in 5 
dimensions and discuss some of its phenomenological aspects.  We begin by 
reviewing the bulk structure of the model of Ref.~\cite{Kawamura:2000ev}.
The 5d spacetime is a direct product of 4d Minkowski spacetime 
$M^4$ and an extra dimension compactified on the $S^1/(Z_2 \times Z_2')$ 
orbifold, with coordinates $x^\mu$ $(\mu = 0,1,2,3)$ and $y$ $(=x^5)$, 
respectively.  The $S^1/(Z_2 \times Z_2')$ orbifold can conveniently be 
viewed as a circle of radius $R$ divided by two $Z_2$ transformations; 
$Z_2$: $y \to -y$ and $Z_2'$: $y' \to -y'$ where $y' = y - \pi R/2$.
The physical space is then an interval $y: [0, \pi R/2]$ which has 
two branes at the two orbifold fixed points at $y=0$ and $\pi R/2$. 
(The branes at $y=\pi R$ and $-\pi R/2$ are identified with those 
at $y=0$ and $\pi R/2$, respectively.)

Under the $Z_2 \times Z_2'$ symmetry, a generic 5d bulk field 
$\phi(x^\mu, y)$ has a definite transformation property
\begin{eqnarray}
  \phi(x^\mu, y) \to \phi(x^\mu, -y) &=& P \phi(x^\mu, y), \\
  \phi(x^\mu, y') \to \phi(x^\mu, -y') &=& P' \phi(x^\mu, y'),
\end{eqnarray}
where the eigenvalues of $P$ and $P'$ must be $\pm 1$.
Denoting the field with $(P, P') = (\pm 1, \pm 1)$ by $\phi_{\pm \pm}$, 
we obtain the following mode expansions \cite{Barbieri:2000vh}: 
\begin{eqnarray}
  \phi_{++} (x^\mu, y) &=& 
      \sum_{n=0}^{\infty} \frac{1}{\sqrt{2^{\delta_{n,0}} \pi R}} \, 
      \phi^{(2n)}_{++}(x^\mu) \cos{2ny \over R},
\label{eq:expansion-1}
\\
  \phi_{+-} (x^\mu, y) &=& 
      \sum_{n=0}^{\infty} \frac{1}{\sqrt{\pi R}} \,
      \phi^{(2n+1)}_{+-}(x^\mu) \cos{(2n+1)y \over R},
\\
  \phi_{-+} (x^\mu, y) &=& 
      \sum_{n=0}^{\infty} \frac{1}{\sqrt{\pi R}} \,
      \phi^{(2n+1)}_{-+}(x^\mu) \sin{(2n+1)y \over R},
\\
  \phi_{--} (x^\mu, y) &=& 
      \sum_{n=0}^{\infty} \frac{1}{\sqrt{\pi R}} \,
      \phi^{(2n+2)}_{--}(x^\mu) \sin{(2n+2)y \over R},
\label{eq:expansion-4}
\end{eqnarray}
where 4d fields $\phi^{(2n)}_{++}$, $\phi^{(2n+1)}_{+-}$, 
$\phi^{(2n+1)}_{-+}$ and $\phi^{(2n+2)}_{--}$ acquire masses 
$2n/R$, $(2n+1)/R$, $(2n+1)/R$ and $(2n+2)/R$ upon compactification.
Zero-modes are contained only in $\phi_{++}$ fields, so that the 
matter content of the massless sector is smaller than that of the 
full 5d multiplet.

In the 5d bulk, we have $SU(5)$ gauge supermultiplets and two Higgs 
hypermultiplets that transform as ${\bf 5}$ and $\bar{\bf 5}$.  
The 5d gauge supermultiplet contains a vector boson 
$A_M$ ($M=0,1,2,3,5$), two gauginos $\lambda$ and $\lambda'$, 
and a real scalar $\sigma$, which is decomposed into a vector 
supermultiplet $V(A_\mu, \lambda)$ and a chiral multiplet in the adjoint 
representation $\Sigma((\sigma+iA_5)/\sqrt{2}, \lambda')$ under $N=1$ 
supersymmetry in 4d.  The hypermultiplet, which consists of two complex 
scalars $\phi$ and $\phi^c$ and two Weyl fermions $\psi$ and 
$\psi^c$, forms two 4d $N=1$ chiral multiplets 
$\Phi(\phi, \psi)$ and $\Phi^c(\phi^c, \psi^c)$ transforming 
as conjugate representations with each other under the gauge group.  
Here $\Phi$ runs over the two Higgs hypermultiplets, $H_{\bf 5}$ and 
$H_{\bar{\bf 5}}$.  ($H_{\bf 5}^c$ and $H_{\bar{\bf 5}}^c$ 
transform as $\bar{\bf 5}$ and ${\bf 5}$ under the $SU(5)$.)

\begin{table}
\begin{center}
\begin{tabular}{|c|c|c|}
\hline
 $(P,P')$ & 4d $N=1$ superfield & mass       \\ \hline
 $(+,+)$  & $V^a$, $H_F$, $H_{\bar{F}}$                  & $2n/R$     \\ 
 $(+,-)$  & $V^{\hat{a}}$, $H_C$, $H_{\bar{C}}$          & $(2n+1)/R$ \\ 
 $(-,+)$  & $\Sigma^{\hat{a}}$, $H_C^c$, $H_{\bar{C}}^c$ & $(2n+1)/R$ \\ 
 $(-,-)$  & $\Sigma^a$, $H_F^c$, $H_{\bar{F}}^c$         & $(2n+2)/R$ \\ 
\hline
\end{tabular}
\end{center}
\caption{The $(Z_2, Z_2')$ transformation properties for the bulk 
gauge and Higgs multiplets.}
\label{ta:Z2-Z2}
\end{table}
The 5d $SU(5)$ gauge symmetry is ``broken'' by the orbifold compactification
to a 4d $SU(3)_C \times SU(2)_L \times U(1)_Y$ gauge symmetry by choosing 
$P=(+,+,+,+,+)$ and $P'=(-,-,-,+,+)$ acting on the ${\bf 5}$. 
Each $Z_2$ reflection is taken to preserve the same 4d $N=1$ supersymmetry.
The $(Z_2, Z_2')$ charges for all components of the vector and Higgs 
multiplets are shown in Table~\ref{ta:Z2-Z2}.
Here, the indices $a$ and $\hat{a}$ denote the unbroken and broken 
$SU(5)$ generators, $T^a$ and $T^{\hat{a}}$, respectively.
The $C$ and $F$ represent the color triplet and weak doublet components 
of the Higgs multiplets, respectively: $H_{\bf 5} \supset \{ H_C, H_F \}$, 
$H_{\bar{\bf 5}} \supset \{ H_{\bar{C}}, H_{\bar{F}} \}$, 
$H_{\bf 5}^c \supset \{ H_C^c, H_F^c \}$ and 
$H_{\bar{\bf 5}}^c \supset \{ H_{\bar{C}}^c, H_{\bar{F}}^c \}$.
Since only $(+,+)$ fields have zero modes, the massless sector consists 
of $N=1$ $SU(3)_C \times SU(2)_L \times U(1)_Y$ vector multiplets 
$V^{a(0)}$ with two Higgs doublet chiral superfields $H_F^{(0)}$ and 
$H_{\bar{F}}^{(0)}$. 
Thus, the doublet-triplet splitting problem is naturally solved in 
this framework \cite{Kawamura:2000ev}.  The higher modes for the vector 
multiplets $V^{a(2n)}$ $(n>0)$ eat $\Sigma^{a(2n)}$ becoming massive 
vector multiplets, and similarly for the $V^{\hat{a}(2n+1)}$ and 
$\Sigma^{\hat{a}(2n+1)}$ $(n \geq 0)$.  An interesting point is that 
the non-zero modes for the Higgs fields have mass terms of the form 
$H_F^{(2n)} H_F^{c(2n)}$, $H_{\bar{F}}^{(2n)} H_{\bar{F}}^{c(2n)}$, 
$H_C^{(2n+1)} H_C^{c(2n+1)}$ and 
$H_{\bar{C}}^{(2n+1)} H_{\bar{C}}^{c(2n+1)}$, which will become 
important when we consider dimension 5 proton decay later.

How should quarks and leptons be incorporated into this theory? 
In this paper we concentrate on the case where they 
are localized on a brane.\footnote{
An interesting alternative is to put quarks and leptons in the bulk.
Then, if we only introduce hypermultiplets $(T_{\bf 10}, T_{\bf 10}^c)$ 
and $(F_{\bar{\bf 5}}, F_{\bar{\bf 5}}^c)$ in the bulk, the 
low-energy matter content is not that of the minimal supersymmetric 
standard model.  However, if we further introduce 
$(T_{\bf 10}^\prime, T_{\bf 10}^{\prime c})$ 
and $(F_{\bar{\bf 5}}^\prime, F_{\bar{\bf 5}}^{\prime c})$ and assign 
opposite $P'$ parities for $T_{\bf 10}$ and $T_{\bf 10}^\prime$ 
(and for $F_{\bar{\bf 5}}$ and $F_{\bar{\bf 5}}^\prime$), it is 
possible to recover the correct low-energy matter content.  The Yukawa 
couplings are placed on either $y=(0,\pi R)$ or $y=\pm \pi R/2$ branes.
The resulting theory is not ``grand unified theory'' in the usual sense, 
since $D$ and $L$ ($Q$ and $U, E$) come from different (hyper)multiplets.
The proton decay from broken gauge boson exchange is absent, and 
there is no $SU(5)$ relation among the Yukawa couplings.
Nevertheless, the theory still keeps the desired features of the 
``grand unified theory'': the quantization of hypercharges 
and the unification of the three gauge couplings.}
An important point then is that when we 
introduce brane-localized matter and/or interactions on some orbifold 
fixed point $O$, they must preserve only the gauge symmetry 
with non-vanishing gauge transformations at this point $\xi^a_O \neq 
0$. This corresponds to the symmetry which 
remains unbroken on this fixed point after compactification.
Specifically, if we introduce the quark and lepton fields on the branes 
at $y=(0,\pi R)$ their multiplet structures and interactions must preserve
$SU(5)$ symmetry and $N=1$ supersymmetry, while if they are located on the 
$y=\pm \pi R/2$ branes they only have to preserve $N=1$ supersymmetry and 
the standard model gauge symmetry.  
For the $SU(5)$ gauge symmetry to provide an understanding of the
quark and lepton quantum numbers, they must reside
on the $y=(0,\pi R)$ branes.
Therefore, we put three generations of $N=1$ chiral superfields
$T_{\bf 10} \supset \{ Q, U, E \}$ 
and $F_{\bar{\bf 5}} \supset \{ D, L \}$  
on the $y=(0,\pi R)$ branes.

What are the transformation properties of the quark and lepton 
superfields under the $Z_2 \times Z_2'$ symmetry?  The parities $P$ 
under $Z_2$ must obviously be plus.  Then, the parities $P'$ under 
$Z_2'$ are determined by the requirement that any operators written 
on the $y=(0,\pi R)$ branes must transform $SU(5)$ covariantly under 
$Z_2'$.  This is because if various terms in an $SU(5)$-invariant operator 
on the $y=0$ brane transform differently under the $Z_2'$, the 
corresponding terms on the $y=\pi R$ brane are not $SU(5)$ invariant, 
contradicting the observation made in Eq.~(\ref{eq:gauge-transformation}).
In particular, since the kinetic terms for the $T_{\bf 10}$ and 
$F_{\bar{\bf 5}}$ fields must transform covariantly under the 
$Z_2'$, there are only four possibilities for the assignment of the 
$P'$ quantum numbers: (i) $P'(Q,U,D,L,E) = \pm(+,-,-,+,-)$ or (ii) 
$P'(Q,U,D,L,E) = \pm(-,+,-,+,+)$.\footnote{
This assignment is different from that adopted in 
Ref.~\cite{Altarelli:2001qj}.} The case $(-,+,-,+,+)$ follows from 
requiring $(-,-,-,+,+)$ on each ${\bf 5}$ index, and the other three 
possibilities result from overall sign changes on $T_{\bf 10}$ and/or 
$F_{\bar{\bf 5}}$. 

Once the $P'$ quantum numbers for the brane fields are determined as 
above, we can work out the transformation properties of the Yukawa 
coupling $[T_{\bf 10} T_{\bf 10} H_{\bf 5}]_{\theta^2}$ and 
$[T_{\bf 10} F_{\bar{\bf 5}} H_{\bar{\bf 5}}]_{\theta^2}$ 
located on the brane.
It turns out that $P'(T_{\bf 10} T_{\bf 10} H_{\bf 5}) = 
P'(T_{\bf 10} F_{\bar{\bf 5}} H_{\bar{\bf 5}}) = -$ in the case (i), 
and $P'(T_{\bf 10} T_{\bf 10} H_{\bf 5}) = 
-P'(T_{\bf 10} F_{\bar{\bf 5}} H_{\bar{\bf 5}}) = -$ in the case (ii).
Therefore, the $Z_2 \times Z_2'$ invariant Yukawa interactions are 
\begin{eqnarray}
  {\cal L}_5 &=& \int d^2\theta \Biggl[
    \frac{1}{2} \{ \delta(y) - \delta(y-\pi R) \}
    \sqrt{2 \pi R}\, y_u T_{\bf 10} T_{\bf 10} H_{\bf 5} 
\nonumber\\
&& + \frac{1}{2} \{ \delta(y) \mp \delta(y-\pi R) \}
    \sqrt{2 \pi R}\, y_d T_{\bf 10} F_{\bar{\bf 5}} H_{\bar{\bf 5}}
  \Biggr] + {\rm h.c.},
\label{eq:5d-Yukawa}
\end{eqnarray}
where $\mp$ takes $-$ and $+$ in the case of (i) and (ii), 
respectively.
(We can easily check that the above expression is invariant under both 
$Z_2: \{ y \to -y,\, T_{\bf 10} T_{\bf 10} H_{\bf 5} \to 
T_{\bf 10} T_{\bf 10} H_{\bf 5},\, 
T_{\bf 10} F_{\bar{\bf 5}} H_{\bar{\bf 5}} \to 
T_{\bf 10} F_{\bar{\bf 5}} H_{\bar{\bf 5}} \}$ and $Z_2': 
\{ y' \to -y',\, T_{\bf 10} T_{\bf 10} H_{\bf 5} \to 
-T_{\bf 10} T_{\bf 10} H_{\bf 5},\, 
T_{\bf 10} F_{\bar{\bf 5}} H_{\bar{\bf 5}} \to 
\mp T_{\bf 10} F_{\bar{\bf 5}} H_{\bar{\bf 5}} \}$.)
After integrating out the extra dimensional coordinate $y$, we get 
$SU(5)$-invariant Yukawa couplings in 4d:
\begin{eqnarray}
  {\cal L}_4 &=& \sum_{n=0}^{\infty} \int d^2\theta \Biggl[
    \sqrt{2} y_u \left( \frac{1}{\sqrt{2^{\delta_{n,0}}}} Q U H_F^{(2n)} 
    + Q Q H_C^{(2n+1)} + U E H_C^{(2n+1)} \right) 
\nonumber\\
&& + \sqrt{2} y_d \left( 
    \frac{1}{\sqrt{2^{\delta_{n,0}}}} Q D H_{\bar{F}}^{(2n)} 
    + \frac{1}{\sqrt{2^{\delta_{n,0}}}} L E H_{\bar{F}}^{(2n)} 
    + Q L H_{\bar{C}}^{(2n+1)} + U D H_{\bar{C}}^{(2n+1)} \right) \Biggr]
  + {\rm h.c.},
\label{eq:4d-Yukawa}
\end{eqnarray}
where the interactions between the zero-mode Higgs doublets and quarks and 
leptons are precisely those of the minimal supersymmetric $SU(5)$ model.
(Note that the $SU(5)$ transformation caused by the broken generators 
$T^{\hat{a}}$ mixes different levels of KK modes, due to 
non-trivial profiles of the broken gauge transformation parameters 
$\xi^{\hat{a}}(y)$ in the extra dimension.)

\subsection{Gauge coupling unification}

We now discuss some phenomenological issues of the model,
beginning with gauge coupling unification.  Since the heavy unified gauge 
boson masses are given by $1/R$, it is natural to identify 
$1/R = M_U$, the unification scale, and this is indeed correct
to zero-th order approximation.  Then, the corrections to 
this naive identification come from two sources.

First, since the brane interactions only have to preserve symmetries 
which remain unbroken locally at that fixed point, we can introduce 
brane kinetic terms for the $SU(3)_C$, $SU(2)_L$ and $U(1)_Y$ gauge 
fields at $y=\pm \pi R/2$ with three different gauge couplings.  
(The operators on the $y=\pm \pi R/2$ branes do not have to preserve 
the full $SU(5)$, since the broken gauge transformation parameters 
$\xi^{\hat{a}}$ vanish on these branes.)
However, we can expect that the effect of these $SU(5)$-violating brane 
kinetic terms is small by making the following observation.
Consider the 5d theory where the bulk and brane gauge couplings 
have almost equal strength.  Then, after integrating out $y$, the 
zero-mode gauge couplings are dominated by the bulk contributions 
because of the spread of the wavefunction of the zero-mode gauge boson.  
Since the bulk gauge couplings are necessarily $SU(5)$ symmetric, 
we find that the $SU(5)$-violating effect coming from brane ones is 
highly suppressed.

To illustrate the above point explicitly, we here consider an extreme 
case where the 5d theory is truly strongly coupled at the cutoff scale 
$M_*$ where the field theory is presumably incorporated into some more 
fundamental theory.  Then, the bulk and brane gauge couplings $g_5$ 
and $g_{4i}$ defined by
\begin{eqnarray}
  {\cal L}_5 &=& \int d^2\theta \left[ 
    \frac{1}{g_5^2} {\cal W}^{\alpha} {\cal W}_{\alpha} 
  + \frac{1}{2} \left\{ \delta(y-\frac{\pi}{2}R) 
    + \delta(y+\frac{\pi}{2}R) \right\}
    \frac{1}{g_{4i}^2} {\cal W}_i^{\alpha} {\cal W}_{i\alpha} \right]
  + {\rm h.c.},
\end{eqnarray}
are estimated to be $1/g_5^2 \sim M_*/(24 \pi^3)$ and 
$1/g_{4i}^2 \sim 1/(16 \pi^2)$ using a strong coupling analysis in 
higher dimensions \cite{Chacko:2000hg}.  Here, $g_5$ and $g_{4i}$ are the  
$SU(5)$-invariant and $SU(5)$-violating contributions. Note that the brane
contributions vanish for ${\cal W}_X^{\alpha}$, which is odd under 
$Z_2'$, but are non-zero and different for ${\cal W}_i^{\alpha}$, 
where $i$ runs over $SU(3)_C$, $SU(2)_L$ and $U(1)_Y$, so that the 
couplings $g_{4i}$ differ from each other by factors of order unity.
On integrating over $y$, we obtain zero-mode 4d gauge couplings $g_{0i}$ 
at the unification scale 
\begin{eqnarray}
  \frac{1}{g_{0i}^2} 
  = \left( \frac{2\pi R}{g_5^2} + \frac{1}{g_{4i}^2} \right)
    \sim \left( \frac{M_* R}{12 \pi^2} + \frac{1}{16 \pi^2} \right).
\end{eqnarray}
Since we know that $g_{0i} \sim 1$, we must take $M_* R \sim 12 \pi^2$ 
in this strongly coupled case.  
This shows that the $SU(5)$-violating contribution from the 
brane kinetic terms on $y=\pm \pi R/2$ (the second term in the above 
equations) is suppressed by an amount equivalent to a 
loop factor $1/(16 \pi^2)$ relative to the 
dominant $SU(5)$-preserving contribution (the first term). 
Of course, this is a tree-level correction, and the brane contribution
is small relative to the bulk term because of the large volume factor
$2 \pi R M_*$.  Note that in the realistic case the theory 
is not necessarily strongly coupled at the cutoff scale 
($M_* R \lsim 12 \pi^2$) as we will see later.  Even then, however, 
it is true that the brane piece is generically suppressed compared with 
the bulk piece, after integrating out $y$, due to the volume factor $2
\pi R M_*$. 
For comparable values of the dimensionless couplings $g_5^2 M_*$ and
$g_{4i}^2$, this corresponds, even in the case of $M_*R =4$, to much 
less than a 1\% correction to the prediction of the weak mixing
angle. 

The second correction to gauge unification originates from the 
running of the gauge 
couplings above the compactification scale due to the KK modes not
filling degenerate $SU(5)$ multiplets.\footnote{
We thank D.~Smith and N.~Weiner for discussions on this issue.}
From the 4d point of view, 
the zero-mode gauge couplings $g_{0i}$ at the compactification scale 
$M_c$ ($=1/R$) is given by \cite{Dienes:1998vh}
\begin{eqnarray}
  \frac{1}{g_{0i}^2(M_c)} \simeq 
    \frac{1}{g_0^2(M_*)} - \frac{b}{8 \pi^2} (M_* R - 1)
    + \frac{b'_i}{8 \pi^2} \ln(M_* R),
\end{eqnarray}
where $b$ and $b'_i$ are constants of $O(1)$.
A crucial observation is that the coefficient $b$ is $SU(5)$ 
symmetric, ie. $b$ is the same for $SU(3)_C$, $SU(2)_L$ and $U(1)_Y$.
This is because the power-law contributions come from renormalizations 
of 5d kinetic terms, which must be $SU(5)$ symmetric.
Since the sum of the first two terms must be $O(1)$ to have 
$g_{0i}(M_c) \sim 1$, the $SU(5)$-violating contribution from the 
running (the third term in the right-hand side of the above equation) 
is suppressed by a loop factor because the running is only logarithmic.
Furthermore, the differences 
$b'_i- b'_j$ are smaller than the corresponding differences 
of the beta-function coefficients at low energies 
($(b'_3, b'_2, b'_1) = (0, 2, 6)$), so that we 
arrive at the following picture.  As the three gauge couplings are
evolved from the weak scale to higher energies, they approach each other 
with the usual logarithmic running, and at the compactification scale 
$M_c$ they take almost equal values but are still slightly different.  
The further relative evolution above $M_c$ is slower, 
but finally the three couplings unify at the cutoff scale $M_*$, where
our 5d theory is incorporated into some more fundamental theory. Note
that $M_c < M_U < M_*$, where $M_U$ is the usual zero-th order value of
the unification scale, $M_U \simeq 2 \times 10^{16}$ GeV. We assume
the physics above $M_*$ to be $SU(5)$ symmetric.

To estimate the threshold correction coming from this second 
source, we consider the one-loop renormalization group equations for the
three gauge couplings.  Assuming that the couplings take a unified value 
$g_*$ at $M_*$, they take the following form:
\begin{eqnarray}
  \alpha_i^{-1}(m_Z) &=& \alpha_*^{-1}(M_*)
    + \frac{1}{2\pi} \Biggl\{ \alpha_i \ln\frac{m_{\rm SUSY}}{m_Z} 
    + \beta_i \ln\frac{M_*}{m_Z} \nonumber\\
  && \qquad \qquad 
    + \gamma_i \sum_{n=0}^{N_l} \ln\frac{M_*}{(2n+2)M_c}
    + \delta_i \sum_{n=0}^{N_l} \ln\frac{M_*}{(2n+1)M_c} \Biggr\}, 
\label{eq:rge}
\end{eqnarray}
where $(\alpha_1, \alpha_2, \alpha_3) = (-5/2, -25/6, -4)$, 
$(\beta_1, \beta_2, \beta_3) = (33/5, 1, -3)$, 
$(\gamma_1, \gamma_2, \gamma_3) = (6/5, -2, -6)$ and 
$(\delta_1, \delta_2, \delta_3) = (-46/5, -6, -2)$.
Here, we have assumed a common mass $m_{\rm SUSY}$ for the superparticles 
for simplicity, and the sum on $n$ includes all KK modes 
below $M_*$, so that $(2N_l+2)M_c \leq M_*$.\footnote{
Terminating the sum of KK modes at $M_*$ may be justified in more 
fundamental theory such as string theory \cite{Ghilencea:2000cy}.}
Contributions proportional to $\gamma_i$ result from the KK modes of
the Higgs doublets and of the standard model gauge bosons and their
$N=2$ partners, while terms proportional to $\delta_i$ result 
from the KK modes of the Higgs triplets and from the vectors and chiral
adjoints of the broken $SU(5)$ generators.
In 4d unified theories one typically only considers threshold
corrections from a few $SU(5)$-split multiplets with mass close to
$M_U$. In the present theory there are $SU(5)$-split KK multiplets
right up to $M_*$.

Taking a linear combination of the three equations, we obtain
\begin{eqnarray}
  (5 \alpha_1^{-1} - 3 \alpha_2^{-1} - 2 \alpha_3^{-1})(m_Z) 
  = \frac{1}{2\pi} \Biggl\{ 8 \ln\frac{m_{\rm SUSY}}{m_Z} 
  + 36 \ln\frac{(2N_l+2)M_c}{m_Z} 
  - 24 \sum_{n=0}^{N_l} \ln\frac{(2n+2)}{(2n+1)} \Biggr\},
\end{eqnarray}
where we have set $M_* = (2N_l+2)M_c$.
The corresponding linear combination in the usual 4d minimal supersymmetric 
$SU(5)$ grand unified theory takes the form
\begin{eqnarray}
  (5 \alpha_1^{-1} - 3 \alpha_2^{-1} - 2 \alpha_3^{-1})(m_Z) 
  = \frac{1}{2\pi} \Biggl\{ 8 \ln\frac{m_{\rm SUSY}}{m_Z} 
  + 36 \ln\frac{M_U}{m_Z} \Biggr\},
\end{eqnarray}
where $M_U = (M_\Sigma^2 M_V)^{1/3}$, and
$M_\Sigma$ and $M_V$ are the adjoint Higgs and the broken gauge boson 
masses, respectively \cite{Hisano:1992mh}.  Therefore, we find the 
following correspondence between the two theories:
\begin{eqnarray}
  \ln\frac{M_c}{m_Z} =
    \ln\frac{M_U}{m_Z} 
    + \frac{2}{3} \sum_{n=0}^{N_l} \ln\frac{(2n+2)}{(2n+1)}
    - \ln(2N_l+2),
\label{eq:correspond-1}
\end{eqnarray}
from the gauge running point of view.  Since 
the experimental values of the gauge couplings
restrict $M_U$ to the range
$1 \times 10^{16}~{\rm GeV} \lsim M_U \lsim 3 \times 10^{16}~{\rm GeV}$, 
we can find the range of $M_c$ for a given $N_l$.  
For instance, if we take $N_l=4$ 
($M_* = 10 M_c$), we find that the compactification scale must be 
in the range
\begin{eqnarray}
  3 \times 10^{15}~{\rm GeV} 
  \;\lsim\; M_c \;\lsim\; 
  8 \times 10^{15}~{\rm GeV},
\label{eq:mc}
\end{eqnarray}
which is somewhat lower than the usual 4d unification scale 
$M_U \simeq 2 \times 10^{16}~{\rm GeV}$.\footnote{
If the fundamental scale of gravity is $M_*$,
the 4 dimensional reduced Planck mass, $M_{\rm pl} \simeq 2.4 \times 
10^{18}~{\rm GeV}$, can result if there are additional dimensions 
in which only gravity propagates with a size of the unification scale.}
There is also an independent bound on 
$N_l$ coming from another linear combination 
$(5 \alpha_1^{-1} - 12 \alpha_2^{-1} + 7 \alpha_3^{-1})(m_Z)$, but 
it is rather weak due to the experimental uncertainty of the strong 
coupling constant $\alpha_3(m_Z)$.

The above estimate is not precise, since 
only the leading logarithmic contributions are included. 
A detailed analysis 
of these corrections to gauge coupling unification will be given in 
Ref.~\cite{HNOS}.  Nevertheless, 
the number given in Eq.~(\ref{eq:mc}) is very encouraging in that the 
dimension 6 proton decay, $p \rightarrow e^+ \pi^0$, induced by the
exchange of an $X$ gauge boson, 
may be seen in the near future.  The present 
experimental limit from Super-Kamiokande, 
$\tau_{p \rightarrow e^+ \pi^0} > 1.6 \times 10^{33}~{\rm years}$ 
\cite{Shiozawa:1998si}, is translated into 
$M_c \gsim 5 \times 10^{15}~{\rm GeV}$, remembering that the $X$ 
gauge boson mass is $M_c$ and the coupling of the $X$ gauge boson to 
quarks is $\sqrt{2}$ times larger than that of the standard model 
gauge boson.  
Thus $N_l \geq 4$, corresponding to a hierarchy $M_*/M_c \geq 10$, could
be excluded by an increase in the experimental limit on 
$\tau_{p \rightarrow e^+ \pi^0}$ by a factor of 6.
In fact, if we require $N_l>1$, so that there is some energy interval 
where the theory is described by higher dimensional field theory 
($M_*/M_c \geq 4$), we obtain an upper bound on the compactification 
scale $M_c < 1.4 \times 10^{16}~{\rm GeV}$, which means that 
increasing the experimental limit by a factor of $60$ covers 
the entire parameter space of the model.

\subsection{$U(1)_R$ symmetry}

Knowing that $1/R$ is somewhat smaller than $M_U$, one may worry about 
too fast proton decay caused by an exchange of the colored Higgs multiplet, 
since they couple to quarks and squarks through the interactions of 
Eq.~(\ref{eq:4d-Yukawa}).  However, these dimension 5 proton decay 
operators are not generated in our theory. The mass terms for the colored 
Higgs supermultiplets are
\begin{eqnarray}
  {\cal L}_4 &=& \sum_{n=0}^{\infty} \int d^2\theta 
    \left( \frac{1}{R} H_C^{(2n+1)} H_C^{c(2n+1)} 
    + \frac{1}{R} H_{\bar{C}}^{(2n+1)} H_{\bar{C}}^{c(2n+1)} \right)
  + {\rm h.c.},
\end{eqnarray}
coupling $H$ to $H^c$. The conjugate fields, $H_C^{c(2n+1)}$ and 
$H_{\bar{C}}^{c(2n+1)}$, 
do not couple directly to the quark and lepton superfields because
of the $U(1)_R$ symmetry shown in Table 2 and discussed below. This
symmetry is only broken by small supersymmetry breaking effects, so
we predict that there is no proton decay induced by dimension 5 operators.

In order to be complete, however, we must also forbid the  
brane interactions
\begin{eqnarray}
  {\cal L}_5 &=& \int d^2\theta \left[
    \frac{1}{2} \{ \delta(y) + \delta(y-\pi R) \}
    H_{\bf 5} H_{\bar{\bf 5}} 
  + \frac{1}{2} \{ \delta(y) \pm \delta(y-\pi R) \}
    T_{\bf 10} T_{\bf 10} T_{\bf 10} F_{\bar{\bf 5}} 
  \right] + {\rm h.c.},
\label{eq:tree-brane}
\end{eqnarray}
where $\pm$ takes $+$ and $-$ in the case of (i) and (ii), respectively.
For this purpose, it is useful to notice that the 5d bulk Lagrangian 
possesses a continuous $U(1)_R$ symmetry.  This bulk $U(1)_R$ is a linear 
combination of the $U(1)$ subgroup of the $SU(2)_R$ automorphism group of 
$N=2$ supersymmetry algebra and a vector-like non-$R$ $U(1)$ symmetry under 
which the Higgs fields transform as $H_{\bf 5}(-1), H_{\bar{\bf 5}}(-1), 
H_{\bf 5}^c(+1), H_{\bar{\bf 5}}^c(+1)$.  The point is that we can 
extend this $U(1)_R$ to the full theory by assigning appropriate charges 
to the quark and lepton superfields.  The full $U(1)_R$ symmetry, with
charge assignments given in Table~\ref{ta:R-charge}, allows the Yukawa 
couplings in Eq.~(\ref{eq:5d-Yukawa}).
\begin{table}
\begin{center}
\begin{tabular}{|c|ccccc|ccc|}  \hline 
  & $\Sigma$ & $H_{\bf 5}$ & $H_{\bar{\bf 5}}$
  & $H_{\bf 5}^c$ & $H_{\bar{\bf 5}}^c$ 
  & $T_{\bf 10}$ & $F_{\bar{\bf 5}}$ & $N_{\bf 1}$ 
\\ \hline
  $U(1)_R$ & 0 & 0 & 0 & 2 & 2 & 1 & 1 & 1 \\ \hline
\end{tabular}
\end{center}
\caption{$U(1)_R$ charges for chiral superfields.}
\label{ta:R-charge}
\end{table}
Hence, imposing the matter quantum numbers of Table~\ref{ta:R-charge},
all interactions 
of Eq.~(\ref{eq:tree-brane}) are forbidden by the $U(1)_R$ symmetry.
This not only solves the problem of dimension 5 proton decay completely 
(both from tree-level brane operators in Eq.~(\ref{eq:tree-brane}) 
and from colored Higgsino exchanges), but also naturally explains why 
the weak Higgs doublets $H_F$ and $H_{\bar{F}}$ are light.
Note that this $U(1)_R$ also forbids unwanted brane operators 
$[F_{\bar{\bf 5}} H_{\bf 5}]_{\theta^2}$ and 
$[T_{\bf 10} F_{\bar{\bf 5}} F_{\bar{\bf 5}}]_{\theta^2}$, 
since it contains the usual $R$-parity as a discrete subgroup.\footnote{
The above $U(1)_R$ allows the brane operator 
$[H_{\bf 5} H_{\bar{\bf 5}}]_{\theta^2 \bar{\theta}^2}$ 
on the $y = (0, \pi R)$ branes.
Thus, it is possible to produce a $\mu$ term (a supersymmetric mass term 
for the Higgs doublets) of the order of the weak scale by the 
mechanism of Ref.~\cite{Giudice:1988yz} after the breakdown of  
$N=1$ supersymmetry, through a constant term in the superpotential 
needed to cancel a positive cosmological constant arising from the 
supersymmetry breaking.  (The superpotential constant term comes from 
an explicit or spontaneous breaking of the $U(1)_R$.)}

\subsection{Supersymmetry breaking}

An important question is the origin of 4 dimensional $N=1$ supersymmetry
breaking. In 4d it is well-known that the supersymmetry breaking must
be isolated to some degree from the particles and interactions of the
minimal supersymmetric standard model. In fact, the argument of
Dimopoulos and Georgi shows that this isolation is only necessary for
the matter fields \cite{Georgi:1974sy}. It is therefore quite clear
that the ideal location for supersymmetry breaking is the branes at 
$y=\pm \pi R/2$. Remarkably, one immediately finds that this results in a
supersymmetry breaking scheme which is precisely that of
the gaugino mediation mechanism \cite{Kaplan:2000ac}. 
Since the gauginos and Higgs fields are in the bulk they have direct
couplings to the supersymmetry breaking field $S$, which we take to be
a gauge and $U(1)_R$ singlet, on the $y=\pm \pi R/2$ branes
\begin{equation}
  {\cal L}_5 =
    \frac{1}{2} \{ \delta(y - \pi R/2) + \delta(y + \pi R/2) \}
  \left[  \int d^2\theta S {\cal W}_i^{\alpha} {\cal W}_{i\alpha} +  
  \int d^4\theta (S^\dagger H_F H_{\bar{F}} + 
  S^\dagger S H_F H_{\bar{F}}) + {\rm h.c.} \right],
\label{eq:gm}
\end{equation}
where coefficients of order unity, in units of $M_*$, are omitted.
Note that $F$-component expectation value for the $S$ field breaks the 
$U(1)_R$ symmetry to $R$ parity.
This generates gaugino masses as well as the $\mu$ and $\mu B$
parameters, while the squarks and sleptons obtain masses 
through radiative corrections so that the supersymmetric flavor problem 
is naturally solved. Superparticle phenomenology is more tightly
constrained than in a general theory of gaugino mediation. The distance
between the supersymmetry breaking and matter branes is known quite
precisely, and, furthermore, the proton decay constraint does not
allow very large values for $M_*/M_c$, so that the ratio of $\mu B/
\mu$ is not a significant problem.
It is also remarkable that, even though we have a unified gauge symmetry
with precise gauge coupling unification, there is no unification of the 
three gaugino masses. They originate from the only location in the theory 
where the gauge transformations $\xi_X(y)$ are forced to vanish, so
that the coefficients of $S {\cal W}_i^{\alpha} {\cal W}_{i\alpha}$ in 
Eq.~(\ref{eq:gm}) is different for the $SU(3)_C, SU(2)_L$ and 
$U(1)_Y$ terms.

\subsection{Quark and lepton masses}

The $SU(5)$ invariance of Eq.~(\ref{eq:5d-Yukawa}) 
guarantees the successful $m_b/m_\tau$ 
mass relation \cite{Chanowitz:1977ye}, but is not realistic, 
since it yields $m_s/m_d = m_\mu/m_e$. 
As discussed at the end of the previous section, if the 
unified symmetry is broken by an orbifold compactification, there is a 
new, constrained mechanism for obtaining $SU(5)$ breaking in fermion 
mass relations. Here we give a simple specific realization of this 
mechanism which is sufficient to allow realistic fermion masses. We add 
bulk hypermultiplets which transform as ${\bf 5} + \bar{\bf 5}$:
$(B,B^c) + (\bar{B}, \bar{B}^c)$.  We assume that they have no 
bulk mass term, for simplicity.  We assign both $B$ and $\bar{B}$ a 
$U(1)_R$ quantum number of +1, so that these hypermultiplets 
should be thought of as matter fields rather than Higgs fields. 
The $Z_2 \times Z_2'$ quantum numbers of these 
hypermultiplets can be chosen such that the zero modes are either weak 
doublets or color triplets; both possibilities lead to realistic 
theories. The $SU(5)$ invariant brane interactions at $y=0$ 
relevant for down-type quark and charged lepton mass matrices are
\begin{equation}
  {\cal L}_5 = \int d^2\theta \delta(y) 
  \left( B(\bar{B} + F_{\bar{\bf 5}}) 
  + T_{\bf 10} (\bar{B} + F_{\bar{\bf 5}}) H_{\bar{\bf 5}} \right),
\label{eq:mattermixing}
\end{equation}
where coupling parameters are omitted. Corresponding interactions are 
placed at $y= \pi R$ to maintain the $Z_2 \times Z_2'$ invariance of the 
theory. The lepton doublets and right-handed down quarks are found to lie 
partly in $\bar{B}$ and partly in $F_{\bar{5}}$, but in different 
combinations. Hence in general the $SU(5)$ relations between the down and 
charged lepton masses are removed. The analysis is very simple when the 
mass terms of Eq.~(\ref{eq:mattermixing}) are smaller than $M_c$. Suppose 
that the zero modes of $B$ are weak doublets. In this case the lepton 
doublet lies partly in $\bar{B}$ and partly in $F_{\bar{\bf 5}}$, while the 
down quark lies completely in $F_{\bar{\bf 5}}$. Hence $T_{\bf 10}
F_{\bar{\bf 5}} H_{\bar{\bf 5}}$ contributes to both down and lepton 
masses, while $T_{\bf 10} \bar{B} H_{\bar{\bf 5}}$ contributes only 
to the lepton masses.  One can imagine that, for some flavor symmetry 
reason, these mixing terms are only important for the 
2-2 entry of the Yukawa matrices, so that the Georgi-Jarlskog mass 
matrices follow \cite{Georgi:1979df}. 

Note that the above mechanism could, in principle, introduce 
supersymmetric flavor problem, since $B$ and $\bar{B}$ fields 
have supersymmetry breaking masses by coupling to $S$ on the 
$y=\pm \pi R/2$ branes and give non-universal squark and slepton 
masses through the mixing with $F_{\bar{\bf 5}}$.  However, the 
amount of induced flavor violation strongly depends on the structure 
of the Yukawa couplings.  Suppose the Yukawa coupling of the 
$\bar{B}$ field, $[T_{\bf 10} \bar{B} H_{\bar{\bf 5}}]_{\theta^2}$, 
is of order one.  Then, the mixing between $\bar{B}$ and the 
second generation $F_{\bar{\bf 5}}$ with angle of $O(0.01)$ will be 
sufficient to break unwanted $SU(5)$ relations on fermion masses.  
In this case, the flavor violating squark and slepton masses are 
suppressed compared with the gaugino masses, and the induced flavor 
violating processes at low energy would be sufficiently small.

Neutrino masses can easily be incorporated into our theory.  
Introducing three right-handed 
neutrino superfields $N_{\bf 1}$'s, with $U(1)_R$ charge of $+1$, on the 
$y=(0, \pi R)$ branes, we can write neutrino Yukawa couplings 
$[F_{\bar{\bf 5}} N_{\bf 1} H_{\bf 5}]_{\theta^2}$ and Majorana 
mass terms for $N_{\bf 1}$'s on the branes, accommodating the 
conventional see-saw mechanism \cite{Seesaw} to explain the smallness 
of the neutrino masses.  Alternatively, we could introduce $N_{\bf 1}$'s 
on the $y=\pm \pi R/2$ branes and forbid their mass terms by imposing 
some symmetry such as $U(1)_{B-L}$.  Then, with appropriate couplings 
to heavy bulk fields of masses $\sim M_*$, exponentially suppressed 
Yukawa couplings, $\exp(-\pi R M_*/2) [F_{\bar{\bf 5}} N_{\bf 1} 
H_{\bf 5}]_{\theta^2}$, are generated via exchanges of these heavy bulk 
fields.  This provides a mechanism of naturally producing small Dirac 
neutrino masses in the present framework.

\section{Conclusions}
\label{section:conc}

It is well-known that compactifying a higher-dimensional gauge 
field theory on a compact manifold leads to gauge symmetry breaking. 
For example, compactification on a circle leads to a mass for all 
gauge bosons corresponding to non-trivial gauge transformations on 
the circle. The only gauge bosons which remain massless  correspond 
to zero-mode gauge transformations.
Compactifying on an orbifold, with an orbifold symmetry which acts 
non-trivially on the gauge bosons, constrains the form of the gauge 
transformations, and reduces the number of their zero modes. This 
removal of zero-mode gauge transformations therefore decreases the 
gauge symmetry of the resulting 4d theory.  Such a reduction can also 
be seen  because the orbifold symmetry removes some of the zero-mode 
states that would be necessary to realize the 4d gauge symmetry.
Kawamura \cite{Kawamura:2000ev} has shown that a 5d theory with 
$SU(5)$ gauge symmetry 
compactified on the orbifold $S^1/(Z_2 \times Z_2')$ elegantly reduces 
the 4d gauge symmetry to $SU(3)_C \times SU(2)_L \times U(1)_Y$, and 
that this is accompanied by a reduction in the zero-mode Higgs states 
to only those that are weak doublets.

We have constructed a complete 5-dimensional $SU(5)$ unified 
field theory compactified on the orbifold $S^1/(Z_2 \times Z_2')$. 
The theory possesses the following features:
\begin{itemize}
\item Quarks and leptons are introduced at orbifold fixed points which 
preserve $SU(5)$ invariance, thereby yielding an understanding of their 
gauge quantum numbers.
\item Quark and lepton masses arise from $SU(5)$ invariant Yukawa couplings 
at these fixed points, which can therefore display $SU(5)$ fermion mass 
relations such as $m_b = m_\tau$.  Mixing between heavy bulk matter and 
brane matter can lead to non-$SU(5)$ symmetric mass relations, as the 
zero-mode structure of the bulk matter is $SU(5)$ violating.
\item The mass matrix for the color triplet Higgs particles is determined 
by compactification, and results in the complete absence of dimension 5 
proton decay from the exchange of these states.
\item Until 4d supersymmetry is broken, the theory possesses an exact 
$U(1)_R$ symmetry. This forbids all proton decay from 
dimension 4 and dimension 5 operators.
\item The orbifold has fixed points where $SU(5)$ is broken. This leads 
to corrections to gauge coupling unification arising from non-$SU(5)$ 
symmetric brane gauge kinetic terms and from KK modes of gauge and Higgs 
multiplets which do not fill degenerate $SU(5)$ multiplets. 
We have argued that the former are negligibly small. 
The latter yield small corrections to the 
weak mixing angle prediction, in a direction which improves the agreement 
between supersymmetric unification and experiment.
\item The $X$ gauge boson, which induces $p \rightarrow e^+ \pi^0$, has a 
mass equal to the compactification scale, $1/R$, which is smaller than 
the usual 4d unification scale of $2 \times 10^{16}$ GeV by a factor 
between 1.4 and 4. We predict that $p \rightarrow e^+ \pi^0$ will be 
discovered by further running of the Super-Kamiokande experiment, or 
at a next generation megaton proton decay detector. 
\item Supersymmetry breaking occurs on a brane distant from the matter 
brane, resulting in the generation of gaugino masses and the $\mu/\mu B$ 
parameters. This breaking is communicated to matter by gaugino mediation, 
so that there is no supersymmetric flavor problem. While supersymmetry 
breaking breaks $U(1)_R$, it preserves $R$ parity. Since the full $SU(5)$ 
gauge transformations do not act on the supersymmetry breaking brane, the 
gaugino mass parameters do not unify.
\item See-saw neutrino masses arise on the matter brane.
\end{itemize}

\section*{Acknowledgements}

We would like to thank N.~Arkani-Hamed, R.~Barbieri, D.E.~Kaplan, 
T.~Okui, D.~Smith and N.~Weiner for valuable discussions.
Y.N. thanks the Miller Institute for Basic Research in Science 
for financial support.
This work was supported by the E.C. under the RTN contract 
HPRn-CT-2000-00148, the Department of Energy under contract 
DE-AC03-76SF00098 and the National Science Foundation under 
contract PHY-95-14797.


\begin{thebibliography}{99}

\bibitem{Georgi:1974yf}
H.~Georgi, H.~R.~Quinn and S.~Weinberg,
Phys.\ Rev.\ Lett.\ {\bf 33}, 451 (1974); \\
S.~Dimopoulos, S.~Raby and F.~Wilczek,
Phys.\ Rev.\ D {\bf 24}, 1681 (1981).

\bibitem{Georgi:1974sy}
H.~Georgi and S.~L.~Glashow,
Phys.\ Rev.\ Lett.\ {\bf 32}, 438 (1974); \\
S.~Dimopoulos and H.~Georgi,
Nucl.\ Phys.\ B {\bf 193}, 150 (1981); \\
N.~Sakai,
Z.\ Phys.\ C {\bf 11}, 153 (1981).

\bibitem{Candelas:1985en}
P.~Candelas, G.~T.~Horowitz, A.~Strominger and E.~Witten,
Nucl.\ Phys.\ B {\bf 258}, 46 (1985); \\
E.~Witten,
Nucl.\ Phys.\ B {\bf 258}, 75 (1985).

\bibitem{Horava:1996qa}
P.~Horava and E.~Witten,
Nucl.\ Phys.\ B {\bf 460}, 506 (1996)
[hep-th/9510209];
Nucl.\ Phys.\ B {\bf 475}, 94 (1996)
[hep-th/9603142].

\bibitem{Kawamura:2000ev}
Y.~Kawamura,
hep-ph/0012125.

\bibitem{Barbieri:2000vh}
R.~Barbieri, L.~J.~Hall and Y.~Nomura,
Phys.\ Rev.\ D {\bf 63}, 105007 (2001)
[hep-ph/0011311].

\bibitem{Altarelli:2001qj}
G.~Altarelli and F.~Feruglio,
hep-ph/0102301.

\bibitem{Sakai:1982pk}
N.~Sakai and T.~Yanagida,
Nucl.\ Phys.\ B {\bf 197}, 533 (1982); \\
S.~Weinberg,
Phys.\ Rev.\ D {\bf 26}, 287 (1982).

\bibitem{Kaplan:2000ac}
D.~E.~Kaplan, G.~D.~Kribs and M.~Schmaltz,
Phys.\ Rev.\ D {\bf 62}, 035010 (2000)
[hep-ph/9911293]; \\
Z.~Chacko, M.~A.~Luty, A.~E.~Nelson and E.~Ponton,
JHEP{\bf 0001}, 003 (2000)
[hep-ph/9911323].

\bibitem{Kawamura:2000nj}
Y.~Kawamura,
Prog.\ Theor.\ Phys.\ {\bf 103}, 613 (2000)
[hep-ph/9902423].

\bibitem{Ellis:1979fg}
J.~Ellis and M.~K.~Gaillard,
Phys.\ Lett.\ B {\bf 88}, 315 (1979).

\bibitem{Georgi:1979df}
H.~Georgi and C.~Jarlskog,
Phys.\ Lett.\ B {\bf 86}, 297 (1979).

\bibitem{Dimopoulos:1992yz}
S.~Dimopoulos, L.~J.~Hall and S.~Raby,
Phys.\ Rev.\ Lett.\ {\bf 68}, 1984 (1992);
Phys.\ Rev.\ D {\bf 45}, 4192 (1992).

\bibitem{Chacko:2000hg}
Z.~Chacko, M.~A.~Luty and E.~Ponton,
JHEP{\bf 0007}, 036 (2000)
[hep-ph/9909248].

\bibitem{Dienes:1998vh}
K.~R.~Dienes, E.~Dudas and T.~Gherghetta,
Phys.\ Lett.\ B {\bf 436}, 55 (1998)
[hep-ph/9803466];
Nucl.\ Phys.\ B {\bf 537}, 47 (1999)
[hep-ph/9806292].

\bibitem{Ghilencea:2000cy}
D.~Ghilencea and G.~Ross,
Nucl.\ Phys.\ B {\bf 569}, 391 (2000)
[hep-ph/9908369].

\bibitem{Hisano:1992mh}
J.~Hisano, H.~Murayama and T.~Yanagida,
Phys.\ Rev.\ Lett.\ {\bf 69}, 1014 (1992).

\bibitem{HNOS}
L.~Hall, Y.~Nomura, T.~Okui and D.~Smith, in preparation. \\
See also, 
Y.~Nomura, D.~Smith and N.~Weiner,
hep-ph/0104041.

\bibitem{Shiozawa:1998si}
M.~Shiozawa {\it et al.}  [Super-Kamiokande Collaboration],
Phys.\ Rev.\ Lett.\ {\bf 81}, 3319 (1998)
[hep-ex/9806014].

\bibitem{Giudice:1988yz}
G.~F.~Giudice and A.~Masiero,
Phys.\ Lett.\ B {\bf 206}, 480 (1988).

\bibitem{Chanowitz:1977ye}
M.~S.~Chanowitz, J.~Ellis and M.~K.~Gaillard,
Nucl.\ Phys.\ B {\bf 128}, 506 (1977).

\bibitem{Seesaw}
T.~Yanagida, 
in {\it Proc. of the Workshop on the Unified Theory and 
Baryon Number in the Universe}, 
ed. O.~Sawada and A.~Sugamoto 
(KEK report 79-18, 1979), p. 95; \\
M.~Gell-Mann, P.~Ramond, and R.~Slansky, 
in {\it Supergravity}, 
ed. P.~van Nieuwenhuizen and D.Z.~Freedman 
(North Holland, Amsterdam, 1979), p. 315.

\end{thebibliography}
\end{document}